\documentclass{iau}
\usepackage{graphicx}

\title[JD 11.~~SCExAO First Results] 
{SCExAO: First Results and On-Sky Performance}

\author[Currie, Guyon, and Martinache et al. ]   
{Thayne Currie$^1$, Olivier Guyon$^{2}$, Frantz Martinache$^{2}$, 
Christophe Clergeon$^{2}$, Michael McElwain$^{3}$, Christian Thalmann$^{4}$, 
Nemanja Jovanovic$^{2}$, Garima Singh$^{2}$, Tomoyuki Kudo$^{2}$}

\affiliation{$^1$Department of Astronomy and Astrophysics, University of Toronto\\
email: {\tt currie@astro.utoronto.ca}\\
[\affilskip]
$^2$National Astronomical Observatory of Japan\\
[\affilskip]
$^3$NASA-Goddard Space Flight Center\\
[\affilskip]
$^4$Astronomical Institute "Anton Pannekoek," University of Amsterdam}

\pubyear{2013}
\volume{299}  
\pagerange{xxxx}
\jname{Exploring the Formation and Evolution of Planetary Systems}
\editors{B. Matthews \& J. Graham, eds.}
\begin{document}

\maketitle
\begin{abstract}
We present new on-sky results for the Subaru Coronagraphic Extreme 
Adaptive Optics imager (SCExAO) verifying and quantifying the contrast gain 
enabled by key components: the closed-loop coronagraphic low-order wavefront 
sensor (CLOWFS) and focal plane wavefront control (``speckle nulling"). 
SCExAO will soon be coupled with a high-order, Pyramid 
wavefront sensor which will yield $>$ 90\% Strehl ratio and enable
 10$^{6}$--10$^{7}$ contrast at small angular separations 
allowing us to image gas giant planets at solar system scales.  
Upcoming instruments like VAMPIRES, FIRST, and CHARIS
will expand SCExAO's science capabilities.
\end{abstract}
\firstsection 
\section{Introduction}
SCExAO is a next-generation extreme-AO platform designed to image 
and characterize the spectra of self-luminous jovian planets 
at solar system-like scales around nearby, young stars (\cite[Martinache and Guyon 2009]{Martinache2009}).  
Components comprising SCExAO are being commissioned in 
several phases.  Already in Phase 1 we have verified the performance of the \textit{Phase Induced Amplitude 
Apodization} (PIAA) coronagraph, which improves our ability to image planets at $\approx$ 
2 $\lambda$/D and with full efficiency by better suppressing diffraction at the telescope pupil edge 
(\cite[Martinache et al. 2012a]{Martinache2012a}). 
Here, we describe on-sky verification of
other key ``Phase 1" components: the CLOWFS (\cite[Guyon et al. 2011]{Guyon2011}),
which corrects for tip-tilt errors to yield $\approx$ 10$^{-3}$ $\lambda$/D pointing accuracy,  
and ``speckle nulling", which uses a deformable mirror (DM) to cancel out static 
speckles limiting planet detections at small separations.  Finally, we 
describe the timeline for implementing SCExAO's high-order wavefront control and new instrumentation.

\section{Phase 1 Engineering Observations}
On November 14, 2012, we observed Pollux with SCExAO combined with the facility 
adaptive optics system (AO-188) and HiCIAO to test and verify the on-sky performance of the CLOWFS and 
speckle nulling components of SCExAO.
After a close-loop sequence of speckle nulling, using images acquired by the SCExAO 
internal science camera (XS), we obtained a sequence of images acquired with HiCIAO alternating 
between the DM flat volt-map and the DM speckle nulling volt-map (e.g. \cite[Martinache et al. 2012b]{Martinache2012b}).   

\begin{figure}[h]
\begin{center}
\includegraphics[width=3.75in,height=1.25in,trim=3mm 7mm 0mm 7mm,clip]{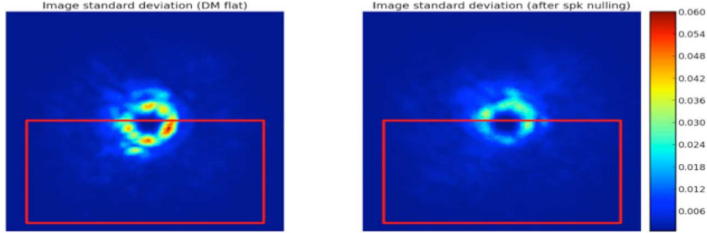}
 \vspace*{-0.35 cm}
 \caption{\textit{SCExAO noise map of the combined image without (left) and with speckle nulling.}}
   \label{fig1}
\end{center}
\end{figure}
Figure \ref{fig1} displays the combined image obtained without (left) and with (right) 
speckle nulling.  Despite the XS's high read noise, the speckle nulling algorithm successfully (blindly) calibrated 
the brightest diffraction features of the long exposure image.  For individual images, contrast gain is moderate 
($\sim$ 2$\times$ for the brightest part of the first ring), but real and persistent.
Speckle nulling reduces the combined image's standard deviation over the entire frame by $\sim$ one-half.
Inside the control region (boxed region), speckle nulling lowers the image standard 
deviation by about a factor of $\sim$ 7.  Thus, at the smallest angular separations (2-4 $\lambda$/D),
speckle nulling alone improves our detection limits by $\approx$ 1-2 magnitudes (Martinache et al. 2013, in prep.).  
\begin{figure}[h]
 \vspace*{-.35 cm}
\begin{center}
\includegraphics[width=5.6in,height=4.7cm,trim=5mm 10mm 0mm 25mm,clip]{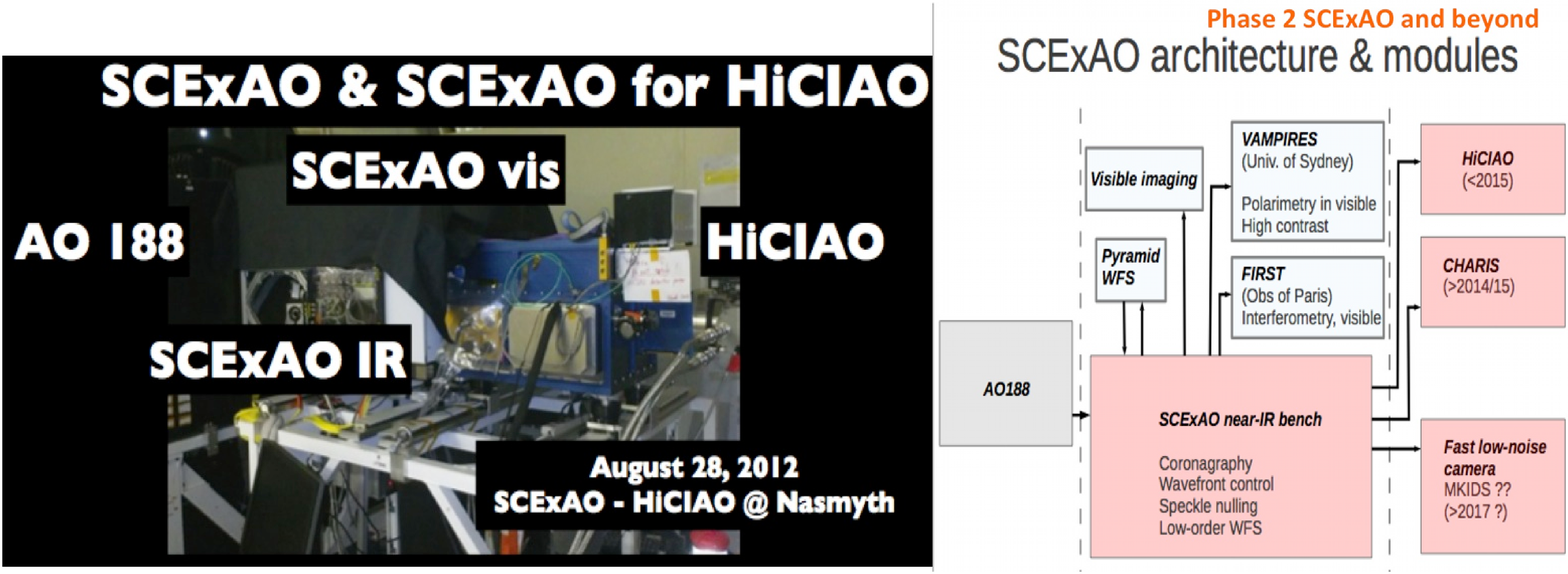}
 \vspace*{-.5 cm}
 \caption{\textit{SCExAO as configured in Phase 1 (left) and Phase 2 (and beyond) (right).}}
   \label{fig2}
\end{center}
\vspace*{-1.0cm}
\end{figure}
\section{Current Timeline and Future Instrumentation}
\textit{SCExAO} has now passed its Phase 1 requirements.  Phase 2 of SCExAO commissioning in Summer/Fall 2013 
will incorporate the 2000-actuator DM controlled by a high-order Pyramid wavefront sensor, which will yield $>$ 90\% Strehl 
at 1.6 $\mu m$ (Figure \ref{fig2}).  Combined with \textit{PIAA}; \textit{CLOWFS}; ``speckle nulling" 
calibrated from a faster, lower noise internal camera; and \textit{angular differential imaging} 
plus aggressive post-processing (\cite[Marois et al. 2006]{Marois2006}, \cite[Currie et al. 2012]{Currie2012}); 
SCExAO should then achieve 10$^{6}$--10$^{7}$ contrasts at $\approx$ 0.1--0.5" and thus image numerous young jovian planets 
at solar system-like scales.  Interferometric instruments VAMPIRES and FIRST capable of studying planet 
formation at even smaller scales will be commissioned at the same time.  Finally, CHARIS 
will eventually replace HiCIAO as the primary SCExAO high-contrast instrument and yield
 near-IR spectra of young planets (\cite[McElwain et al. 2013]{McElwain2013}) to probe the planets' atmospheres.

\end{document}